\documentclass[12pt,a4paper]{article}

\usepackage{ifthen} 
\newboolean{pdflatex}
\setboolean{pdflatex}{true} 

\newboolean{articletitles}
\setboolean{articletitles}{true} 

\newboolean{uprightparticles}
\setboolean{uprightparticles}{false} 

\newboolean{inbibliography}
\setboolean{inbibliography}{false} 

\usepackage{microtype}
\usepackage{lineno}  
\usepackage{xspace} 
\usepackage{caption} 

\usepackage{graphicx}  
\usepackage{color}
\usepackage{colortbl}
\graphicspath{{./figs/}} 

\usepackage{amsmath} 
\usepackage{amssymb}
\usepackage{amsfonts}
\usepackage{upgreek} 

\newcommand*\patchAmsMathEnvironmentForLineno[1]{%
\expandafter\let\csname old#1\expandafter\endcsname\csname #1\endcsname
\expandafter\let\csname oldend#1\expandafter\endcsname\csname
end#1\endcsname
 \renewenvironment{#1}%
   {\linenomath\csname old#1\endcsname}%
   {\csname oldend#1\endcsname\endlinenomath}%
}
\newcommand*\patchBothAmsMathEnvironmentsForLineno[1]{%
  \patchAmsMathEnvironmentForLineno{#1}%
  \patchAmsMathEnvironmentForLineno{#1*}%
}
\AtBeginDocument{%
\patchBothAmsMathEnvironmentsForLineno{equation}%
\patchBothAmsMathEnvironmentsForLineno{align}%
\patchBothAmsMathEnvironmentsForLineno{flalign}%
\patchBothAmsMathEnvironmentsForLineno{alignat}%
\patchBothAmsMathEnvironmentsForLineno{gather}%
\patchBothAmsMathEnvironmentsForLineno{multline}%
\patchBothAmsMathEnvironmentsForLineno{eqnarray}%
}

\usepackage{hyperref}    
\usepackage[all]{hypcap} 

\def\MagUp {\mbox{\em Mag\kern -0.05em Up}\xspace}

\ifthenelse{\boolean{uprightparticles}}%
{

 \def\Peta        {\ensuremath{\upeta}\xspace}

 \def\Ppi         {\ensuremath{\uppi}\xspace}

 \def\PDelta      {\ensuremath{\Delta}\xspace}                 
 \def\PXi      {\ensuremath{\Xi}\xspace}                 
 \def\PLambda      {\ensuremath{\Lambda}\xspace}                 
 \def\PSigma      {\ensuremath{\Sigma}\xspace}                 
 \def\POmega      {\ensuremath{\Omega}\xspace}                 
 \def\PUpsilon      {\ensuremath{\Upsilon}\xspace}                 
 
 \def\PB      {\ensuremath{\mathrm{B}}\xspace}                 
                  
 \def\PD      {\ensuremath{\mathrm{D}}\xspace}

 \def\PK      {\ensuremath{\mathrm{K}}\xspace}

 \def\Pi      {\ensuremath{\mathrm{i}}\xspace}

 \def\Ps      {\ensuremath{\mathrm{s}}\xspace}

}
{

 \def\Peta        {\ensuremath{\eta}\xspace}

 \def\Ppi         {\ensuremath{\pi}\xspace}

 \mathchardef\PDelta="7101
 \mathchardef\PXi="7104
 \mathchardef\PLambda="7103
 \mathchardef\PSigma="7106
 \mathchardef\POmega="710A
 \mathchardef\PUpsilon="7107
                  
 \def\PB      {\ensuremath{B}\xspace}                 
                  
 \def\PD      {\ensuremath{D}\xspace}

 \def\PK      {\ensuremath{K}\xspace}

 \def\Pi      {\ensuremath{i}\xspace}

 \def\Ps      {\ensuremath{s}\xspace}

}

\makeatletter
\ifcase \@ptsize \relax
  \newcommand{\miniscule}{\@setfontsize\miniscule{4}{5}}
\or
  \newcommand{\miniscule}{\@setfontsize\miniscule{5}{6}}
\or
  \newcommand{\miniscule}{\@setfontsize\miniscule{5}{6}}
\fi
\makeatother

\DeclareRobustCommand{\optbar}[1]{\shortstack{{\miniscule (\rule[.5ex]{1.25em}{.18mm})}
  \\ [-.7ex] $#1$}}

\def\squark    {{\ensuremath{\Ps}}\xspace}

\def\pion   {{\ensuremath{\Ppi}}\xspace}
\def\piz    {{\ensuremath{\pion^0}}\xspace}

\def\pip    {{\ensuremath{\pion^+}}\xspace}
\def\pim    {{\ensuremath{\pion^-}}\xspace}
\def\pipm   {{\ensuremath{\pion^\pm}}\xspace}
\def\pimp   {{\ensuremath{\pion^\mp}}\xspace}

\def\kaon    {{\ensuremath{\PK}}\xspace}
  \def\Kbar    {{\kern 0.2em\overline{\kern -0.2em \PK}{}}\xspace}

\def\KorKbar    {\kern 0.18em\optbar{\kern -0.18em K}{}\xspace}
\def\Kz      {{\ensuremath{\kaon^0}}\xspace}

\def\Kp      {{\ensuremath{\kaon^+}}\xspace}
\def\Km      {{\ensuremath{\kaon^-}}\xspace}
\def\Kpm     {{\ensuremath{\kaon^\pm}}\xspace}
\def\Kmp     {{\ensuremath{\kaon^\mp}}\xspace}
\def\KS      {{\ensuremath{\kaon^0_{\rm\scriptscriptstyle S}}}\xspace}
\def\KL      {{\ensuremath{\kaon^0_{\rm\scriptscriptstyle L}}}\xspace}
\def\KSorL   {{\ensuremath{\kaon^0_{\rm\scriptscriptstyle S,L}}}\xspace}

\newcommand{\etapr}{\ensuremath{\Peta^{\prime}}\xspace}

  \def\Dbar    {{\kern 0.2em\overline{\kern -0.2em \PD}{}}\xspace}
\def\D       {{\ensuremath{\PD}}\xspace}
\def\Db      {{\ensuremath{\Dbar}}\xspace}
\def\DorDbar    {\kern 0.18em\optbar{\kern -0.18em D}{}\xspace}
\def\Dz      {{\ensuremath{\D^0}}\xspace}
\def\Dzb     {{\ensuremath{\Dbar{}^0}}\xspace}

\def\B       {{\ensuremath{\PB}}\xspace}
\def\Bbar    {{\ensuremath{\kern 0.18em\overline{\kern -0.18em \PB}{}}}\xspace}

\def\BorBbar    {\kern 0.18em\optbar{\kern -0.18em B}{}\xspace}
\def\Bz      {{\ensuremath{\B^0}}\xspace}
\def\Bzb     {{\ensuremath{\Bbar{}^0}}\xspace}

\def\Bs      {{\ensuremath{\B^0_\squark}}\xspace}

  \def\Y#1S{\ensuremath{\PUpsilon{(#1S)}}\xspace}

\def\Lbar        {{\ensuremath{\kern 0.1em\overline{\kern -0.1em\PLambda}}}\xspace}
\def\LorLbar    {\kern 0.18em\optbar{\kern -0.18em \PLambda}{}\xspace}

\def\to                 {\ensuremath{\rightarrow}\xspace}

\def\CP                {{\ensuremath{C\!P}}\xspace}

\def\AT#1     {\ensuremath{A_{\mathrm{T}}^{#1}}\xspace}           

\def\C#1      {\ensuremath{\mathcal{C}_{#1}}\xspace}                       
\def\Cp#1     {\ensuremath{\mathcal{C}_{#1}^{'}}\xspace}                    
\def\Ceff#1   {\ensuremath{\mathcal{C}_{#1}^{\mathrm{(eff)}}}\xspace}        
\def\Cpeff#1  {\ensuremath{\mathcal{C}_{#1}^{'\mathrm{(eff)}}}\xspace}       
\def\Ope#1    {\ensuremath{\mathcal{O}_{#1}}\xspace}                       
\def\Opep#1   {\ensuremath{\mathcal{O}_{#1}^{'}}\xspace}                    

\newcommand{\tev}{\ifthenelse{\boolean{inbibliography}}{\ensuremath{~T\kern -0.05em eV}\xspace}{\ensuremath{\mathrm{\,Te\kern -0.1em V}}}\xspace}
\newcommand{\gev}{\ensuremath{\mathrm{\,Ge\kern -0.1em V}}\xspace}
\newcommand{\mev}{\ensuremath{\mathrm{\,Me\kern -0.1em V}}\xspace}
\newcommand{\kev}{\ensuremath{\mathrm{\,ke\kern -0.1em V}}\xspace}
\newcommand{\ev}{\ensuremath{\mathrm{\,e\kern -0.1em V}}\xspace}
\newcommand{\gevc}{\ensuremath{{\mathrm{\,Ge\kern -0.1em V\!/}c}}\xspace}
\newcommand{\mevc}{\ensuremath{{\mathrm{\,Me\kern -0.1em V\!/}c}}\xspace}
\newcommand{\gevcc}{\ensuremath{{\mathrm{\,Ge\kern -0.1em V\!/}c^2}}\xspace}
\newcommand{\gevgevcccc}{\ensuremath{{\mathrm{\,Ge\kern -0.1em V^2\!/}c^4}}\xspace}
\newcommand{\mevcc}{\ensuremath{{\mathrm{\,Me\kern -0.1em V\!/}c^2}}\xspace}

\def\gsim{{~\raise.15em\hbox{$>$}\kern-.85em
          \lower.35em\hbox{$\sim$}~}\xspace}
\def\lsim{{~\raise.15em\hbox{$<$}\kern-.85em
          \lower.35em\hbox{$\sim$}~}\xspace}

\def\tell1  {TELL1\xspace}
\def\ukl1   {UKL1\xspace}

\newcommand{\eg}{\mbox{\itshape e.g.}\xspace}
\newcommand{\ie}{\mbox{\itshape i.e.}\xspace}

\newcommand{\etc}{\mbox{\itshape etc.}\xspace}

\usepackage{cite} 
\usepackage{mciteplus}

\def\phani   {\phantom{1}}
\def\phanm   {\phantom{-}}
\def\phanid  {\phantom{1.}}
\def\phanii  {\phantom{11}}
\def\phaniid {\phantom{11.}}

\begin{document}

\renewcommand{\thefootnote}{\fnsymbol{footnote}}
\setcounter{footnote}{1}

\begin{titlepage}
\pagenumbering{roman}

{\bf\boldmath\huge
\begin{center}
  Contributions to the width difference in the neutral \D system from hadronic decays
\end{center}
}

\vspace*{2.0cm}

\begin{center}
  T.~Gershon$^1$, J.~Libby$^2$, G.~Wilkinson$^{3,4}$
\bigskip\\
{\it\footnotesize 
$ ^1$ Department of Physics, University of Warwick, Coventry, United Kingdom\\
$ ^2$ Indian Institute of Technology Madras, Chennai 600036, India\\
$ ^3$ University of Oxford, Denys Wilkinson Building, Keble Road, OX1 3RH, United Kingdom \\
$ ^4$ European Organisation for Nuclear Research (CERN), CH-1211, Geneva 23, Switzerland \\
}
\end{center}

\vspace{\fill}

\begin{abstract}
  \noindent
  Recent studies of several multi-body $\Dz$ meson decays have revealed that the final states are dominantly \CP-even.
  However, the small value of the width difference between the two physical eigenstates of the $\Dz$--$\Dzb$ system indicates that the total widths of decays to \CP-even and \CP-odd final states should be the same to within about a percent.
  The known contributions to the width difference from hadronic \Dz decays are discussed, and it is shown that an apparent excess of quasi-\CP-even modes is balanced, within current uncertainty, by interference effects in quasi-flavour-specific decays.
  Decay modes which may significantly affect the picture with improved measurements are considered.
\end{abstract}

\vspace{\fill}

\end{titlepage}

\newpage
\setcounter{page}{2}
\mbox{~}

\cleardoublepage

\renewcommand{\thefootnote}{\arabic{footnote}}
\setcounter{footnote}{0}

\pagestyle{plain} 
\setcounter{page}{1}
\pagenumbering{arabic}

\section{Introduction}
\label{sec:intro}

Oscillations between the flavour eigenstates of the neutral \D meson result in physical states with distinct masses and widths. 
Parameters that quantify the mass and width differences are given by 
\begin{equation}
  \label{eq:yDdef}
  x_D \equiv (m_2 - m_1)/\Gamma_D \qquad {\rm and} \qquad y_D \equiv (\Gamma_2 - \Gamma_1)/(2\Gamma_D) \, ,
\end{equation}
where $m_{1(2)}$ and $\Gamma_{1(2)}$ are the masses and widths of the physical eigenstates 
and $\Gamma_D \equiv (\Gamma_1 + \Gamma_2)/2$ is the average width.
These are now known to good precision from studies of the evolution with decay time of the rates for certain \D decays, the latest world averages~\cite{HFAG} giving
\begin{equation}
  \label{eq:yDHFAG}
  x_D = (4.1\,^{+1.4}_{-1.5})\times 10^{-3} \, , \qquad 
  y_D = (6.3\,^{+0.7}_{-0.8})\times 10^{-3} \, .
\end{equation}

The branching fractions for \Dz decays to \CP-even and \CP-odd final states are related to the reduced width difference $y_D$, in the limit of \CP conservation, by
\begin{equation}
  \label{eq:yDBF}
  \small
  y_D = \frac{\Gamma(\Dz \to \CP{\rm \text -even}) - \Gamma(\Dz \to \CP{\rm \text -odd})}{2\Gamma_D} = \frac{{\cal B}(\Dz \to \CP{\rm \text -even}) - {\cal B}(\Dz \to \CP{\rm \text -odd})}{2} \, .
\end{equation}
In Eq.~(\ref{eq:yDBF}) the widths and branching fractions to \CP-even and \CP-odd states include not only decays to \CP eigenstates, but any decays with net \CP content~\cite{Falk:2001hx}.
This can be quantified by the fractional \CP-even content $F_+$ ($F_+ = 1$ corresponds to pure \CP-even, $F_+=0$ to pure \CP-odd).
Decays with net \CP content include quasi-\CP eigenstates (\ie self-conjugate final states with $F_+ \neq \frac{1}{2}$) and also quasi-flavour-specific decays.
Considering these different types of decays, the relation for $y_D$ can conveniently be expressed 
\begin{equation}
  \label{eq:yDBF2}
  y_D = \frac{1}{2} \sum_i (2 F_{+\,i} -1) {\cal B}(\Dz \to f_i) \, ,
\end{equation}
where the sum is over all distinct decays to hadronic final states $f_i$, each with net \CP content $2 F_{+\,i} -1$.
Semileptonic and other nonhadronic decays do not contribute as they are either assumed to be flavour-specific, hence $2 F_{+} - 1 = 0$, or have branching fractions that are negligibly small.

Measurements of the $F_+$ values of various decay modes have recently become available, from studies of samples of quantum-correlated $\psi(3770) \to \D\Db$ decays.
These include direct measurements as well as determinations of quantities that can be translated to provide information on $F_+$. 
It is therefore of interest, and is the main purpose of this paper, to evaluate Eq.~(\ref{eq:yDBF2}) from experimental data, and to compare the value obtained with that from Eq.~(\ref{eq:yDHFAG}).
Throughout the paper \CP conservation in $\Dz$--$\Dzb$ mixing and decay, which is known to be a good approximation, is assumed.

In addition to the intrinsic value of such a survey, the outcome may be useful to identify additional \D decay modes which could be used to study charm mixing and \CP violation, and also for determinations of the angle $\gamma$ of the Unitarity Triangle formed from elements of the Cabibbo-Kobayashi-Maskawa (CKM) quark mixing matrix~\cite{Cabibbo:1963yz,Kobayashi:1973fv}.
Decay modes with high net \CP content are particularly useful to probe these phenomena.
Many of the \CP eigenstate modes with largest branching fractions, however, give decay topologies (\eg $\KS\piz$) without any charged track originating from the \D decay vertex, and therefore cannot be used to study decay-time-dependent effects.
If decay modes with high net \CP content and experimentally accessible topologies can be identified, they could be used to improve the precision of charm mixing and \CP violation parameters~\cite{Malde:2015xra}.

The CKM angle $\gamma$ can be determined from $B \to DK$ decays with negligible theoretical uncertainty when the neutral \D meson is reconstructed in final states to which both \Dz and \Dzb can decay.
Methods have been proposed, and analyses performed, with \D decays to \CP eigenstates (referred to as the GLW method)~\cite{Gronau:1990ra,Gronau:1991dp}, doubly-Cabibbo-suppressed decays (ADS)~\cite{Atwood:1996ci,Atwood:2000ck} and self-conjugate multibody \Dz decays (GGSZ)~\cite{Giri:2003ty,Bondar,Poluektov:2004mf}.
Ultimately, the best precision on $\gamma$ is obtained by combining results from all of these methods and others~\cite{LHCb-PAPER-2013-020,LHCb-CONF-2013-006,Lees:2013nha,Trabelsi:2013uj}, and it is important to continue to improve the precision in all of them.

In case the final state of a multibody $D$ decay is dominated by a particular \CP-eigenvalue, it can be used in a ``quasi-GLW'' analysis~\cite{Nayak:2014tea}.
First quasi-GLW analyses of $B \to DK$ with $\D \to \pip\pim\piz$ and $\D \to \Kp\Km\piz$ have recently been reported by LHCb~\cite{LHCb-PAPER-2015-014}.
If additional decay modes with high net \CP content can be identified, they could also be used to improve the precision on $\gamma$.
Similarly, multibody quasi-flavour-specific decays can be used in a ``quasi-ADS'' analysis~\cite{Atwood:2000ck,Atwood:2003mj}, as has been done with $\D \to K^\mp \pi^\pm \pi^0$~\cite{Lees:2011up,Nayak:2013tgg,LHCb-PAPER-2015-014} and $\Kpm\pimp\pip\pim$~\cite{LHCb-PAPER-2012-055} decays; in this case the sensitivity depends on a ``coherence factor''~\cite{Atwood:2003mj} that is also related to the net \CP content of the final state, as will be shown.
Therefore modes with high net \CP content are also very useful to improve the sensitivity to $\gamma$.

The remainder of the paper is organised as follows.
In Sec.~\ref{sec:qCP}, the current knowledge of charm decays to \CP eigenstates and (quasi-)\CP eigenstates is reviewed, while Sec.~\ref{sec:qFS} contains a similar discussion of quasi-flavour-specific channels.
In Sec.~\ref{sec:odd}, channels potentially worthy of future experimental investigation are identified.
A brief summary is given in Sec.~\ref{sec:summary}.

\section{\boldmath Charm decays to (quasi-)\CP eigenstates}
\label{sec:qCP}

The current knowledge~\cite{PDG2014} of branching fractions of hadronic $D^0$ decay modes to \CP eigenstates or quasi-\CP eigenstates is given in Table~\ref{tab:big}.
The corresponding values of  $2F_+ - 1$ are also given.  
Summing all these values according to Eq.~(\ref{eq:yDBF2}) gives 
\begin{equation}
  y_D^{{\cal B}-{\rm q}\CP} = \left( 11.0 \pm 1.2 \right) \times 10^{-3} \, ,
\end{equation}
where the notation $y_D^{{\cal B}-{\rm q}\CP}$ indicates that this value is determined from the experimental knowledge of the branching fractions of decays to quasi-\CP eigenstates only.
It is not a true estimate of $y_D$ as there are additional modes, including those discussed in Secs.~\ref{sec:qFS} and~\ref{sec:odd} that are not included in the sum.
Note that the uncertainty on $y_D^{{\cal B}-{\rm q}\CP}$ is dominated by the least precisely measured modes, with the largest contributions coming from uncertainties on the branching fractions of \D decays to $\KS\piz\piz$ and $\KS\eta\piz$ final states.

Table~\ref{tab:big} includes the multibody \D decay modes for which explicit determinations of $F_+$ have been made~\cite{Nayak:2014tea,Malde:2015mha}:
\begin{itemize}
  \setlength{\itemsep}{2pt}
  \setlength{\parskip}{0pt}
  \setlength{\parsep}{0pt}
\item $\Dz \to \pip\pim\piz$: $F_+ = 0.973 \pm 0.017$\,,
\item $\Dz \to \Kp\Km\piz$: $F_+ = 0.732 \pm 0.055$\,,
\item $\Dz \to \pip\pim\pip\pim$: $F_+ = 0.737 \pm 0.028$\,.
\end{itemize}
The dominance of \CP-even in $\Dz \to \pip\pim\piz$ decays is particularly striking. 
It should be noted that this feature has been previously discussed in terms of isospin~\cite{Gaspero:2008rs} and of the impact on the determination of $\gamma$ from Dalitz plot analysis of $B \to DK$, $\D \to \pip\pim\piz$ decays~\cite{Aubert:2007ii}.
A potential explanation in the context of flavour-SU(3) and factorisation has been discussed~\cite{Bhattacharya:2010id}, but there is as-yet no fundamental understanding of the \CP-even dominance in these decays.

\begin{table}[!htb]
  \caption{
    Branching fractions of hadronic $D^0$ decay modes to \CP eigenstates or
    quasi-\CP eigenstates.
    Values are taken from Ref.~\cite{PDG2014} unless otherwise specified.
    The particles $\pi, \kaon, \eta, \etapr$ and $\omega$ are considered as
    stable; decays involving other particles, such as $\rho$ or $\phi$ mesons,
    are accounted for under the corresponding multibody final state.
    Effects of \CP violation in the neutral kaon system are at the level of
    ${\cal O}(10^{-3})$ and are negligible.
  }
  \label{tab:big}
  \centering
  \begin{tabular}{c@{\hspace{10mm}}cc@{\hspace{10mm}}cc}
    \hline
    Decay mode & $2F_+ - 1$ & & ${\cal B}/10^{-3}$ \\
    \hline
    \multicolumn{5}{c}{\CP-even} \\
    $\Kp\Km$ & 1 & & $\phani3.96\phani \pm 0.08\phani$ \\
    $\KS\KS$ & 1 & & $\phani0.17\phani \pm 0.04\phani$ \\
    $\KL\piz$ & 1 & & $10.0\phanii \pm 0.7\phanii$ \\
    $\pip\pim$ & 1 & & $\phani1.402 \pm 0.026$ \\
    $\piz\piz$ & 1 & & $\phani0.820 \pm 0.035$ \\
    $\piz\eta$ & 1 & & $\phani0.68\phani \pm 0.06\phani$ & \cite{PDG2014,Weidenkaff:2015awa} \\
    $\piz\etapr$ & 1 & & $\phani0.90\phani \pm 0.14\phani$ \\
    $\piz\omega$ & 1 & & $\phani0.11\phani \pm 0.04\phani$ & \cite{Weidenkaff:2015awa} \\
    $\eta\eta$ & 1 & & $\phani1.67\phani \pm 0.20\phani$ \\
    $\eta\etapr$ & 1 & & $\phani1.05\phani \pm 0.26\phani$ \\
    $\KS\KS\KS$ & 1 & & $\phani0.91\phani \pm 0.13\phani$ \\
    $\KS\piz\piz$ & 1 & & $\phani9.1\phanii \pm 1.1\phanii$ \\
    $\KS\eta\piz$ & 1 & & $\phani5.5\phanii \pm 1.1\phanii$ \\
    \hline
    \multicolumn{5}{c}{Mostly \CP-even} \\
    $\pip\pim\piz$ & $0.946 \pm 0.034$ & \cite{Malde:2015mha} & $14.3\phanii \pm 0.6\phanii$ \\ 
    $\Kp\Km\piz$ & $0.464 \pm 0.110$ & \cite{Malde:2015mha} & $\phani3.29\phani \pm 0.14\phani$ \\ 
    $\pip\pim\pip\pim$ & $0.474 \pm 0.056$ & \cite{Malde:2015mha} & $\phani7.42\phani \pm 0.21\phani$ \\ 
    \hline
    \multicolumn{5}{c}{Approximately \CP-neutral} \\
    $\KS\pip\pim$ & $0.112 \pm 0.024$ & \cite{Briere:2009aa,Libby:2010nu} & $28.3\phanii \pm 2.0\phanii$ \\
    $\KS\Kp\Km$ & $0.194 \pm 0.064$ & \cite{Briere:2009aa,Libby:2010nu} & $\phani4.60\phani \pm 0.16\phani$ & \cite{PDG2014,Weidenkaff:2015awa} \\
    $\Kp\Km\pip\pim$ & $0.14 \pm 0.21$ & \cite{Artuso:2012df} & $\phani2.43\phani \pm 0.12\phani$ \\ 
    \hline 
    \multicolumn{5}{c}{\CP-odd} \\
    $\KS\piz$ & $-1$ & & $11.9\phanii \pm 0.4\phanii$ \\
    $\KS\eta$ & $-1$ & & $\phani4.79\phani \pm 0.30\phani$ \\
    $\KS\etapr$ & $-1$ & & $\phani9.4\phanii \pm 0.5\phanii$ \\
    $\KS\omega$ & $-1$ & & $11.1\phanii \pm 0.6\phanii$ \\
    $\piz\piz\piz$ & $-1$ & & $<0.4$ \\
    $\KS\KS\piz$ & $-1$ & & $<0.6$ \\
    \hline
  \end{tabular}
\end{table}

For some other multibody \D decay modes, although there is no direct determination of $F_+$, it is possible to obtain constraints from published information. 
For $\Dz \to \KS\pip\pim$ or $\KS\Kp\Km$ decays, this can be done in one of two ways:
(i) from the ratios of double-tagged and single-tagged yields for each \CP-eigenstate;
(ii) from the relation between $F_+$ and the factors $K_i$ and $c_i$~\cite{Nayak:2014tea},
\begin{equation}
  F_+ = \sum_{i>0} \frac{1}{2}\left( K_i + K_{-i} + 2 c_i \sqrt{K_i K_{-i}} \right) \, ,
\end{equation}
where $K_i$ is the fraction of flavour-tagged \D decays that fall into Dalitz plot bin $i$ ($\sum_{i>0} K_i + K_{-i} = 1$), $c_i$ is the decay-rate weighted average of the cosine of the strong phase difference between \Dz and \Dzb decays in bin $i$~\cite{Giri:2003ty,Bondar:2005ki,Bondar:2008hh}, and bins $i$ and $-i$ are related by symmetry under charge conjugation of the final state.
The values quoted in Table~\ref{tab:big} are determined with the second method, which should be more precise.
The same approach can also be used to determine for $\Dz \to \KL\pip\pim$, $2 F_+ - 1 = -0.288 \pm 0.036$ and for $\Dz \to \KL\Kp\Km$, $2 F_+ - 1 = -0.136 \pm 0.088$; since the branching fractions for these modes have not been measured they are not included in Table~\ref{tab:big} (but see discussion below).
For the $\Dz \to \Kp\Km\pip\pim$ mode, only approach (i) is possible; an estimate of $F_+$ could in principle be made from the amplitude model for that decay~\cite{Artuso:2012df}, but the evaluation of uncertainties would be difficult.

The majority of \CP-odd final states are of the form $\KS h^0$ ($h^0 = \piz, \eta, \etapr, \omega$).
For each of these there is a corresponding \CP-even final state, $\KL h^0$.
Among the decays with $\KL$ mesons in the final state, only the branching fraction for $\Dz \to \KL\piz$ has been measured~\cite{He:2007aj}.
The result differs from that for $\Dz \to \KS\piz$ by about $4\tan^2\theta_C \approx 20\,\%$ of its absolute value ($\theta_C$ is the Cabibbo angle) due to interference between the Cabibbo-favoured and doubly-Cabibbo-suppressed amplitudes~\cite{Bigi:1994aw}.
Similar relations are expected between the other $\KSorL h^0$ branching fractions where $h^0$ is a pseudoscalar ($\eta, \etapr$)~\cite{Rosner:2006bw,Bhattacharya:2008ss,Bhattacharya:2009ps}, though somewhat different effects are possible in other cases~\cite{Bhattacharya:2008ke}.
Comparisons of the $2F_+-1$ values for $\KSorL \pip\pim$ and $\KSorL\Kp\Km$ discussed above also indicate approximately opposite \CP content between the final states containing \KS and \KL mesons.

Thus, for many of the \CP-odd final states, there are expected to be corresponding \CP-even final states with branching fractions of comparable magnitude.
The converse is not true: for the majority of \CP-even final states (apart from the $\KL h^0$ states), there is no corresponding \CP-odd final state.
Essentially, this is due to the fact that singly-Cabibbo-suppressed transitions appear to result predominantly in \CP-even final states, while Cabibbo-favoured transitions can give both \CP-even and \CP-odd states.
Exceptions arise from the three-body \CP-even states $\KS\piz\piz$, $\KS\piz\eta$ and $\KS\KS\KS$, for which corresponding \CP-odd states  $\KL\piz\piz$, $\KL\piz\eta$ and $\KL\KS\KS$ exist.
(Although numerically insignificant, it is interesting to note that there is no \CP-odd counterpart to the $\KS\KS$ mode, as the symmetry of the wavefunction prevents $\Dz\to\KS\KL$ decays.)

This raises a subtle issue in the evaluation of Eq.~(\ref{eq:yDBF2}).
While it is valid to sum all modes for which experimental results are available, this could be argued to be biasing, since more data are available for modes containing \KS than \KL mesons.
Considered together, the net \CP content can be expected to cancel to some extent, although the residual contributions may still be important.
For comparison, if all such modes containing $\Kz$ mesons in the final state are excluded from the calculation of $y_D^{{\cal B}-{\rm q}\CP}$, a larger value of $\left( 14.7 \pm 0.5 \right) \times 10^{-3}$ is obtained.

\section{\boldmath Quasi-flavour-specific hadronic charm decays}
\label{sec:qFS}

Decay modes such as $\Dz \to \Kmp\pipm$ have contributions from both Cabibbo-favoured and doubly-Cabibbo-suppressed amplitudes.
The interference between these two amplitudes leads to an asymmetry in the widths for \CP-even and \CP-odd decays~\cite{Gronau:2001nr,Atwood:2002ak,Asner:2005wf},
\begin{equation}
  \label{eq:AcpBES}
  {\cal A}^{\CP}_f = \frac{{\cal B}\left(\D_{\CP-} \to f\right)-{\cal B}\left(\D_{\CP+} \to f\right)}{{\cal B}\left(\D_{\CP-} \to f\right)+{\cal B}\left(\D_{\CP+} \to f\right)} = \frac{y_{\D}-2\, R_f \, r_f \cos \delta_{f}}{1 + R_{{\rm WS}\,f}} \, .
\end{equation}
Here the final state $f$ includes both conjugate states (\eg\ $f = \Kmp\pipm$), $r_f$ is the average ratio of magnitudes of, and $-\delta_f$ is the average strong phase between, the suppressed and favoured amplitudes, $R_f$ is the coherence factor, $0 < R_f < 1$, that quantifies the dilution due to integrating over the phase space (for a two-body decay $r_{f}e^{-i\delta_{f}}$ is the suppressed-to-favoured amplitude ratio and $R_f = 1$), and $R_{{\rm WS}\,f}$ is the decay time integrated ratio of wrong-sign to right-sign rates, including effects from both the doubly-Cabibbo-suppressed amplitude and charm mixing.
Precise definitions of these quantities can be found, for example, in Ref.~\cite{Atwood:2003mj}.
Note that alternative conventions for the definition of $\delta_{f}$ can be found in the literature; in particular, $\delta_{f} \to \delta_{f} + \pi$ is also widely used, \eg in Ref.~\cite{Ablikim:2014gvw}.
Since a positive sign for ${\cal A}^{\CP}_f$ in Eq.~(\ref{eq:AcpBES}) indicates a larger width for \CP-odd decays, neglecting small corrections from $y_{\D}$ and $R_{\rm WS}$, leads to the relation
\begin{equation}
  \label{eq:qFS}
  2F_{+\,f}-1 = -{\cal A}^{\CP}_f = 2\, R_f \, r_f \cos \delta_{f} \, .
\end{equation}

The asymmetry for $\Dz \to \Kmp\pipm$ decays has been measured to be ${\cal A}^{\CP}_{\Kmp\pipm} = (12.7 \pm 1.3 \pm 0.7)\,\%$~\cite{Ablikim:2014gvw} (see also Refs.~\cite{Rosner:2008fq,Asner:2008ft,Asner:2012xb}).
Similar quantities for other quasi-flavour-specific decay modes can also be obtained from published results.
Measurements of the relevant properties of $\Dz \to \Kmp\pipm\piz$ and $\Kmp\pipm\pip\pim$ decays have been performed~\cite{Lowery:2009id,Libby:2014rea}, where the reported quantity $\Delta_{\CP}$ is equivalent to ${\cal A}^{\CP}$.
These values are given in Table~\ref{tab:mid}.
It should be noted that the significantly non-zero value of $\Delta_{\CP}$ for $\Dz \to \Kmp\pipm\pip\pim$ decays is in tension with the small value of the coherence parameter for these decays also reported in Ref.~\cite{Libby:2014rea}.
This arises as the latter uses information from all different tags reconstructed in the $\psi(3770) \to \D\Dbar$ events, while $\Delta_{\CP}$ is evaluated using \CP tags only.
This shows that a more precise determination of $2F_+-1$ for this mode is possible using Eq.~(\ref{eq:AcpBES}) or~(\ref{eq:qFS}) and all available information on the relevant parameters.
Improved measurements with larger $\psi(3770)$ data sets are clearly well motivated.

\begin{table}[!htb]
  \caption{
    Branching fractions of hadronic $D^0$ decay modes to quasi-flavour-specific final states.
    Values are taken from Ref.~\cite{PDG2014} unless otherwise specified.
  }
  \label{tab:mid}
  \centering
  \begin{tabular}{c@{\hspace{10mm}}cc@{\hspace{10mm}}cc}
    \hline
    Decay mode & $2F_+ - 1$ & & ${\cal B}/10^{-3}$ & \\
    \hline
    \multicolumn{5}{c}{Cabibbo-favoured / doubly-Cabibbo-suppressed} \\
    $\Kmp\pipm$ & $-0.127 \pm 0.015$ & \cite{Ablikim:2014gvw} & $\phani38.8 \pm 0.5\phani$ \\
    $\Kmp\pipm\piz$ & $-0.084 \pm 0.028$ & \cite{Libby:2014rea} & $139\phanid \pm 5\phaniid$ \\
    $\Kmp\pipm\pip\pim$ & $-0.119 \pm 0.027$ & \cite{Libby:2014rea} & $\phani80.8 \pm 0.20$ \\
    \hline
    \multicolumn{5}{c}{Singly-Cabibbo-suppressed} \\
    $\KS\Kpm\pimp$ & $\phanm0.65 \pm 0.14$ & \cite{Insler:2012pm} & $\phanii5.6 \pm 0.6\phani$ \\
    \hline
  \end{tabular}
\end{table}

For completeness, it should be noted that there are also singly-Cabibbo-suppressed decays to non-self-conjugate final states, an example of which is $\Dz \to \KS\Kpm\pimp$.
This decay has been studied~\cite{Insler:2012pm} with results that indicate significant net \CP-even content. 
The relation ${\cal A}^{\CP} = \left(\kappa^+ - \kappa^-\right)/\left(\kappa^+ + \kappa^-\right)$, which is valid neglecting charm mixing, can be used to together with the values of $\kappa^\pm$ reported in Ref.~\cite{Insler:2012pm} to obtain ${\cal A}^{\CP}(\KS\Kpm\pimp) = -0.65 \pm 0.14$ (small possible correlations have been neglected).
Alternatively, the right-hand side of Eq.~(\ref{eq:AcpBES}) can be evaluated using quantities determined using all tag information.
(In this case it is not valid to assume $r_f \ll 1$, as done in Eq.~(\ref{eq:qFS}), though charm mixing effects can still be neglected.)
This approach gives a consistent value of ${\cal A}^{\CP}(\KS\Kpm\pimp) = -0.69 \pm 0.08$. 
It would be interesting to see if the dominance of \CP-even in $\Dz \to \KS\Kpm\pimp$ decays is confirmed with studies of larger $\psi(3770)\to\D\Dbar$ decays, or if it is consistent with the recently-obtained amplitude models for these decays~\cite{LHCb-PAPER-2015-026}.
(To obtain an estimate for $\delta$ from amplitude models of $\Dz \to \KS\Kpm\pim$ decays requires knowledge of the relative phase between the amplitude in the two distinct final states.  
This quantity, specifically the phase between the amplitudes for $\Dz \to K^{*+}\Km$ and $K^{*-}\Kp$ decays, can be and has been determined from Dalitz plot analysis of $\Dz \to \Kp\Km\piz$ decays~\cite{Rosner:2003yk,Cawlfield:2006hm,Aubert:2007dc}.)

Summing the values from Table~\ref{tab:mid} according to Eq.~(\ref{eq:yDBF2}) gives $y_D^{{\cal B}-{\rm qFS}} = \left( -11.3 \pm 2.3 \right) \times 10^{-3}$, where the notation $y_D^{{\cal B}-{\rm qFS}}$ indicates that this determination comes from the measured quasi-flavour-specific modes.
If modes with final states containing \Kz mesons are excluded, for the reasons discussed in Sec.~\ref{sec:qCP}, this becomes $\left( -13.1 \pm 2.3 \right) \times 10^{-3}$.
The uncertainty is dominated by the imprecise knowledge of $2F_+-1$ for $\Dz \to \Kmp\pipm\piz$ and $\Kmp\pipm\pip\pim$ decays.
It is striking that all of the $\D \to \Kmp\pipm$, $\Kmp\pipm\piz$ and $\Kmp\pipm\pip\pim$ decays are net \CP-odd.

\section{\boldmath Possible additional modes with net \CP content}
\label{sec:odd}

Using Eq.~(\ref{eq:yDBF2}) and the resulting relation $y_D = y_D^{{\cal B}-{\rm q}\CP} + y_D^{{\cal B}-{\rm qFS}} + y_D^{{\cal B}-{\rm missing}}$ allows a prediction of the net \CP of the decay modes that have not been accounted for, due to absence of experimental information, 
\begin{equation}
  \label{eq:yDBF-miss}
  y_D^{{\cal B}-{\rm missing}} = \frac{1}{2} \sum_{i\, ({\rm missing})} (2 F_{+,\,i} -1) {\cal B}(\Dz \to f_i) = \, \left( 6.6 \pm 2.6 \right) \times 10^{-3} \,,
\end{equation}
or $\left( 4.7 \pm 2.3 \right) \times 10^{-3}$ if final states containing \Kz mesons are excluded.
Since, the Particle Data Group reports that $(38.2\pm1.4)\,\%$ of \Dz decays are as-yet unaccounted for, this indicates that these missing modes either have low values of $2F_+-1$, or that there is a balance between net \CP-even and \CP-odd modes leading to the low value of Eq.~(\ref{eq:yDBF-miss}).
(For comparison, the modes included in Tables~\ref{tab:big} and~\ref{tab:mid} account for about $40\,\%$ of the total \Dz width -- the remainder are predominantly from semileptonic decays.)

\begin{table}[!htb]
  \caption{
    Hadronic $D^0$ decay modes that could be of interest with regard to their \CP content. 
    The branching fractions are given where known~\cite{PDG2014}.
    The branching fraction for $\Dz \to \KSorL \pip\pim\piz$ ($\Kmp\pipm\pip\pim\piz$) includes contributions from $\KSorL\eta$ and $\KSorL\omega$ ($\Kmp\pipm\eta$ and $\Kmp\pipm\omega$) in the final state.
  }
  \label{tab:small}
  \centering
  \begin{tabular}{c@{\hspace{10mm}}c@{\hspace{10mm}}c}
    \hline
    Decay mode & $2F_+ - 1$ & ${\cal B}/10^{-3}$ \\
    \hline
    $\eta \pip\pim$ & unknown & $\phani1.09 \pm 0.16$ \\
    $\eta \piz\piz$ & $-1$ & unknown \\
    $\eta \Kp\Km$ & unknown & unknown \\
    $\eta \KS\KS$ & $-1$ & unknown \\
    $\etapr \pip\pim$ & unknown & $\phani0.45 \pm 0.17$ \\
    $\etapr \piz\piz$ & $-1$ & unknown \\
    \hline
    $\KL\pip\pim$ & $-0.288 \pm 0.036$ & unknown \\
    $\KL\Kp\Km$ & $-0.136 \pm 0.088$ & unknown \\
    \hline
    $\pip\pim\piz\piz$ & unknown & $10.0\phani \pm 0.9\phani$ \\
    $\Kp\Km\piz\piz$ & unknown & unknown \\
    $\KS \pip\pim\piz$ & unknown & $52\phantom{.00} \pm 6\phantom{.00}$ \\
    $\KL \pip\pim\piz$ & unknown & unknown \\
    \hline
    $\Kmp\pipm\piz\piz$ & unknown & unknown \\
    $\Kmp\pipm\pip\pim\piz$ & unknown & $42\phantom{.00} \pm 4\phantom{.00}$ \\
    $\Kmp\pipm\pip\pim\pip\pim$ & unknown & $22\phantom{.00} \pm 6\phantom{.00}$ \\
    \hline
  \end{tabular}
\end{table}

Treating only $\pi, \kaon, \eta, \etapr$ and $\omega$ as stable particles, it is easy to see in Table~\ref{tab:big} that almost all two-body \CP eigenstate modes have been measured.
There are no results for the decays $\Dz \to \omega \eta$ and $\omega \etapr$, but these are unlikely to have large branching fractions.
For three-body decays, in the case that the final state is composed of neutral spin-zero particles ($\KSorL, \piz, \eta, \etapr$), a pure \CP eigenstate is obtained~\cite{Gershon:2004tk}.
Several such decay modes are included in Table~\ref{tab:big}, and there are some others, such as $\eta\piz\piz$ (\CP-odd) that are worthy of investigation.
However, most are unlikely to make a significant contribution to the net \CP content of \Dz decays due the cancellation of modes containing $\KSorL$ mesons and due to the small branching fractions expected for decays with little phase space.
Thus, it is unlikely that pure \CP eigenstates contribute significantly to Eq.~(\ref{eq:yDBF-miss}).

In the case of \Dz decays to $h^0 h^{\prime +} h^{\prime -}$ ($h^0 = \KSorL, \piz, \eta, \etapr$; $h^\prime = \pi, \kaon$),
the final state is in general a mixture of \CP-even and \CP-odd.
Specifically, if the angular momentum between the $h^0$ and the $h^{\prime +} h^{\prime -}$ system is $L$ then the \CP content is $\eta_{h^0}\times(-1)^{L}$, where $\eta_{h^0}$ is the \CP eigenvalue of $h^0$.
Hence if both even and odd values of $L$ contribute then the final state is not a pure \CP eigenstate.
By conservation of angular momentum, $L$ also gives the partial wave in the $h^{\prime +}h^{\prime -}$ system ($L=0 \leftrightarrow$ S~wave; $L=1 \leftrightarrow$ P~wave; \etc).
As discussed above, in the case of $\pip\pim\piz$, the $\pip\pim$ system is dominantly in odd partial waves, while for $\KS\pip\pim$ or $\KS\Kp\Km$ both even and odd waves contribute approximately equally.
For the \Dz decays to $\eta \pip\pim$ and $\etapr \pip\pim$ it is possible that either even or odd partial waves dominate, giving them high net \CP content.
The dominance of even (odd) partial waves is to be expected when the $\pip\pim$ system is mainly isospin zero (one), however if the decay is dominated by contributions such as $\Dz \to a_0^\pm \pimp \to \eta \pipm\pimp$ then the net \CP content appears hard to predict.
In the case, however, of $a_0$ dominance, a sizable branching fraction for $\Dz \to a_0^0 \piz \to \eta \piz\piz$ (\CP-odd) may be expected.

These modes are certainly worth experimental investigation, as is the as-yet unmeasured decay $\Dz \to \eta\Kp\Km$.
However, as the branching fractions of the $\Dz \to \eta \pip\pim$ and $\etapr \pip\pim$~\cite{Artuso:2008aa} decays are ${\cal O}(10^{-3})$ they cannot make large contributions to the total \CP content of \Dz decays.

As discussed in Sec.~\ref{sec:qCP}, there is a lack of experimental results on final states containing \KL mesons.
As the branching fractions for the $\Kz \pip\pim$ and $\Kz\Kp\Km$ decays are relatively high, the residual net \CP content between the final states containing \KS and \KL mesons may be numerically significant.  
In addition to determination of the branching fractions, further investigation of the differences in Dalitz plot distributions between the two states with different physical kaon states may provide insight into hadronic effects in the multibody decays.

As the multiplicity of the final state increases, the na\"ive expectation is that the net \CP content is less likely to be significantly non-zero, as this would require some form of coherence across a multi-dimensional phase space.
However, as discussed above the final state of the $\Dz \to \pip\pim\pip\pim$ decay has been shown to be mainly \CP-even, so it is possible that there are other four- or more-body final states with net \CP content.
The final states $\pip\pim\piz\piz$ and $\Kp\Km\piz\piz$ appear to be worth investigation; the former is known to have a sizable branching fraction~\cite{Rubin:2005py}.
Similarly to the case discussed above, the \CP content is given by the angular momentum of the $h^{\prime +} h^{\prime -}$ system, but in this case even (odd) partial waves correspond to \CP-even (-odd).

Another four-body \Dz decay that is of interest with regard to its \CP content is that to $\KS \pip\pim\piz$.
The contribution from $\KS\omega$ (\CP-odd) comprises only around $20\,\%$ of the total branching fraction, and there may be regions of phase space that are dominated by one or the other \CP eigenvalue.
Although any net \CP content would be balanced, to some extent, from an opposite effect in $\Dz \to \KL\pip\pim\piz$ decays, there may be a non-negligible residual contribution.

It is likely that a large proportion of the unaccounted-for decay modes are from high-multiplicity quasi-flavour-specific hadronic channels such as $\Kmp\pipm\piz\piz$, $\Kmp\pipm\pip\pim\piz$, $\Kmp\pipm\eta$, $\Kmp\pipm\pip\pim\pip\pim$, and so on.
The net \CP-content of such modes is expected to be small, as the coherence factors should be low for the reasons discussed above.
A great deal of additional experimental investigation would be necessary to make a comprehensive survey.

Table~\ref{tab:small} contains a selection of modes for which it is not yet possible to evaluate the contribution to the reduced width difference. 
Experimental investigations of these channels would be of great interest.

\section{Summary}
\label{sec:summary}

There is now a significant body of results from studies of various hadronic \D meson decays, including determinations of net \CP content, coherence factors and parameters related to strong phase differences.
These allow a survey of the contributions to the reduced width difference $y_D$ in the $\Dz$--$\Dzb$ system from different types of hadronic decays.
If only \CP-eigenstate and quasi-\CP-eigenstates are considered, there appears to be an excess of \CP-even modes compared to expectation.
This demonstrates that the pedagogical argument that the width difference arises from such effects is too simplistic.
The effects of interference between doubly-Cabibbo-suppressed and Cabibbo-favoured amplitudes in quasi-flavour-specific modes are found to be equally important; the small value of $y_D$ is related to cancellation between the contributions.
It would be interesting to see if a similar situation occurs in the $\Bz$--$\Bzb$ system, where a small reduced width difference is expected in the Standard Model~\cite{Lenz:2011ti}, but enhancements due to new physics effects are possible~\cite{Dighe:2007gt,Gershon:2010wx,Bobeth:2014rda}.
Existing measurements of hadronic \Bz decays are, however, not complete enough to allow a data-based study like that presented here; even fewer measurements are available in the \Bs system.

It has previously been noted that SU(3) symmetry breaking effects can account for $y_D \sim 1\,\%$ in an ``exclusive'' approach where $D$ decays are assumed to be dominated by a small number of processes~\cite{Falk:2001hx}.
In this approach, the contribution to $y_D$ from the SU(3) multiplet of decays to final states composed of two pseudoscalars is small, but larger contributions are expected from pseudoscalar-vector, vector-vector and multibody decays as larger SU(3)-breaking effects are induced by phase space considerations.
Indeed, it was noted that the contribution to $y_D$ from the U-spin doublet of charged $K$ and $\pi$,
\begin{equation}
  y_{K\pi} = \frac{1}{2} \left( {\cal B}(\Dz \to \pip\pim) + {\cal B}(\Dz\to\Kp\Km) + 2 \, r_{K\pi} \, \cos \delta_{K\pi} \, {\cal B}(\Dz \to \Km\pip) \right) \, ,
\end{equation}
has almost perfect cancellation; with the benefit of current data it is seen that $y_{K\pi} = (0.22 \pm 0.30) \times 10^{-3}$.
It is curious to note that of similar relations for modes with an additional $\piz$ meson or $\pip\pim$ pair,
\begin{eqnarray}
  y_{K\pi\piz} & = & \frac{1}{2} \sum_{i \,\in\, \pip\pim\piz,\,\Kp\Km\piz,\,\Kpm\pimp\piz} (2 F_{+\,i} -1) {\cal B}(\Dz \to f_i) \, , \\
  y_{K\pi\pip\pim} & = & \frac{1}{2} \sum_{i \,\in\, \pip\pim\pip\pim,\,\Kp\Km\pip\pim,\,\Kpm\pimp\pip\pim} (2 F_{+\,i} -1) {\cal B}(\Dz \to f_i) \, ,
\end{eqnarray}
the former is also consistent with zero within current experimental uncertainties ($y_{K\pi\piz} = (1.7 \pm 2.0) \times 10^{-3}$) while the latter shows a small deviation ($y_{K\pi\pip\pim} = (-2.8 \pm 1.1) \times 10^{-3}$).
This may indicate some underlying symmetry of the dynamics of the multibody decays, and also appears to support the hypothesis that $y_D$ arises largely from SU(3)-breaking induced by phase space effects in multibody hadronic \D decays.

Although it is remarkable that existing data on hadronic charm decays allow a
survey such as that presented in this paper, further experimental investigation would be necessary for a full understanding.
Improved measurements of the properties of several channels are well motivated, including those modes discussed in Sec.~\ref{sec:odd} as well as $\Dz \to \Kmp\pipm\piz$ and $\Kmp\pipm\pip\pim$.
It will also be important to reduce the fraction of unaccounted-for \Dz decay modes, which are likely to include some high-multiplicity hadronic channels.
Notably, the BESIII experiment has the potential to improve significantly existing measurements, and to add new results, in particular through exploitation of its data sample of $\psi(3770) \to \Dz\Dzb$ decays.
This sample not only allows challenging decay topologies to be reconstructed but also enables measurements that rely on quantum-correlations to be performed.
Complementary information on hadronic charm decays can also be obtained from other experiments such as Belle~(II) and LHCb; in addition to branching fraction measurements, novel use can be made of charm mixing to determine $F_+$ for quasi-\CP eigenstates~\cite{Malde:2015xra}, and coherence factors and strong phase differences for quasi-flavour-specific modes~\cite{Harnew:2013wea,Harnew:2014zla}.

\section*{Acknowledgements}

The authors wish to thank their former collaborators on the CLEOc experiment, who produced many of the results discussed in this work, and also acknowledge helpful discussions with Marco Gersabeck, Olli Lupton, Alexey Petrov and Mike Sokoloff.
This work is supported by 
the Science and Technology Facilities Council (United Kingdom),
the UK-India Education and Research Initiative, 
and by the European Research Council under FP7.
TG thanks the Munich Institute for Astro- and Particle Physics for hospitality during the period that this work was completed.

\addcontentsline{toc}{section}{References}
\setboolean{inbibliography}{true}
\bibliographystyle{LHCb}
\bibliography{references,main,LHCb-PAPER,LHCb-CONF,LHCb-DP,LHCb-TDR}

\end{document}